\documentclass{article} 
\usepackage{iclr2026_conference,times}

\usepackage{amsmath,amsfonts,bm}

\def\eqref#1{equation~\ref{#1}}

\def\1{\bm{1}}

\DeclareMathAlphabet{\mathsfit}{\encodingdefault}{\sfdefault}{m}{sl}
\SetMathAlphabet{\mathsfit}{bold}{\encodingdefault}{\sfdefault}{bx}{n}

\usepackage{hyperref}
\usepackage{url}
\usepackage{multirow}
\usepackage{booktabs}
\usepackage{float} 
\usepackage{graphicx}
\usepackage[skip=5pt]{caption}

\usepackage{fontawesome5}
\usepackage{xcolor}
\usepackage{twemojis}
\usepackage[T1]{fontenc}

\newcommand{\best}[1]{\textbf{\underline{#1}}}

\iclrfinalcopy 
\begin{document}

\begin{center}
    \vspace*{-3.5em}

    {\LARGE \textbf{GLM-TTS Technical Report}}

    \vspace{0.6em}

    {\normalsize \bfseries
    Jiayan Cui$^{\star 1}$ \quad
    Zhihan Yang$^{\star 1}$ \quad
    Naihan Li$^1$ \quad
    Jiankun Tian$^1$ \quad
    Xingyu Ma$^1$ \\[2pt]
    Yi Zhang$^1$ \quad
    Guangyu Chen$^1$ \quad
    Runxuan Yang$^{1 2}$ \quad
    Zijian Huang$^1$ \quad
    Yuqing Cheng$^1$ \\[2pt]
    Yizhi Zhou$^1$ \quad
    Guochen Yu$^{\dag 1}$ \quad
    Xiaotao Gu$^1$ \quad
    Jie Tang$^2$
    }

    \vspace{0.4em}

{\normalsize
    $^1$Zhipu AI \quad $^2$Tsinghua University \\[2pt]
    $^\star$Equal contribution $\quad$ $^\dag$Project leader
}

    \vspace{0.5em}

    \scriptsize
    \renewcommand{\arraystretch}{1.0}
    \begin{tabular}{@{}l@{\hspace{2mm}}l}
        \faGithub\ \textbf{Code:} & \url{github.com/zai-org/GLM-TTS} \\
        \includegraphics[height=1.8ex]{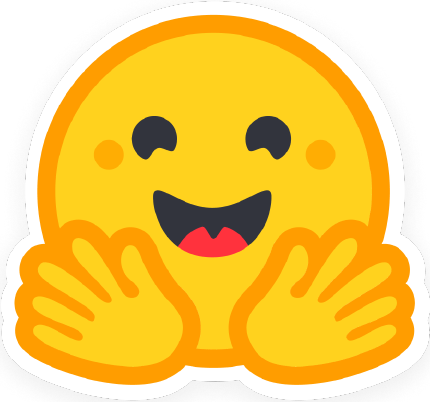} \textbf{Model:} & \url{huggingface.co/zai-org/GLM-TTS} \\
        \faMicrophone\ \textbf{Demo:} & \url{audio.z.ai/} \\
    \end{tabular}
\end{center}

\begin{abstract}
\vspace{-1.0em}
This work proposes GLM-TTS, a production-level TTS system designed for efficiency, controllability, and high-fidelity speech generation. GLM-TTS follows a two-stage architecture, consisting of a text-to-token autoregressive model and a token-to-waveform diffusion model. With only 100k hours of training data, GLM-TTS achieves state-of-the-art performance on multiple open-source benchmarks.
To meet production requirements, GLM-TTS improves speech quality through an optimized speech tokenizer with fundamental frequency constraints and a GRPO-based multi-reward reinforcement learning framework that jointly optimizes pronunciation, speaker similarity, and expressive prosody. In parallel, the system enables efficient and controllable deployment via parameter-efficient LoRA-based voice customization and a hybrid phoneme–text input scheme that provides precise pronunciation control.
Our code is available at \url{https://github.com/zai-org/GLM-TTS}. Real-time speech synthesis demos are provided via Z.ai (\url{audio.z.ai}), the Zhipu Qingyan app/web (\url{chatglm.cn}). 

\end{abstract}

\begin{figure}[!ht]
\begin{center}
\includegraphics[width=.7\textwidth]{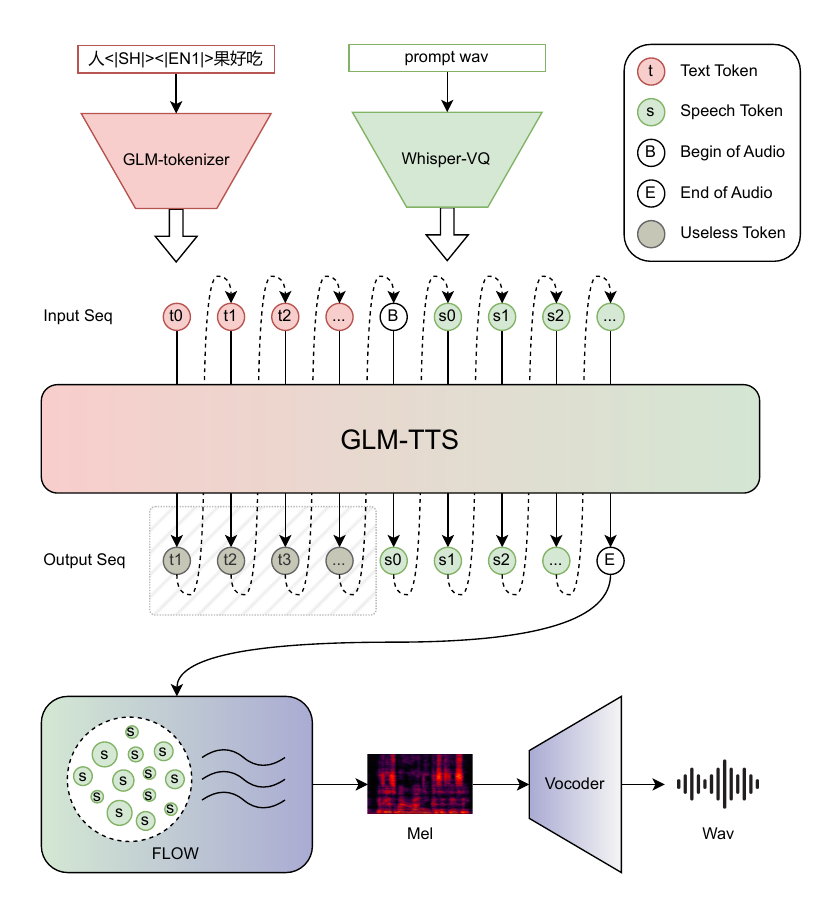}
\end{center}
\caption{Overall Architecture of GLM-TTS}
\label{img:GLM_TTS_overall}
\end{figure}

\section{Introduction}

Text-to-Speech (TTS) synthesis has evolved into a cornerstone technology powering human-computer interaction, content creation, and accessibility tools, from virtual assistants and podcast production to educational narration and dubbing. Over the past decade, TTS systems have evolved from early neural acoustic models that directly predict continuous speech representations—such as attention-based sequence-to-sequence models (e.g., Tacotron~\citep{wang2017tacotronendtoendspeechsynthesis}, Transformer-TTS~\citep{li2019neuralspeechsynthesistransformer}) and non-autoregressive feed-forward architectures (e.g., Fastspeech~\citep{ren2019fastspeechfastrobustcontrollable} series) to Transformer-based~\citep{vaswani2023attentionneed} large language model (LLM)-driven paradigms that treat speech generation as discrete token modeling. This transition has enabled substantial progress in in-context learning (ICL) zero-shot voice cloning, multilingual synthesis, and prosodic naturalness.

Recent advanced TTS models can be broadly categorized into three paradigms: (1) auto-regressive (AR) zero-shot TTS based on neural codec such as SoundStream~\citep{zeghidour2021soundstreamendtoendneuralaudio} and EnCodec~\citep{défossez2022highfidelityneuralaudio} (e.g., VALL-E~\citep{wang2023neural}, Spark-TTS~\citep{wang2025sparkttsefficientllmbasedtexttospeech}); (2) flow-matching or diffusion-based non-autoregressive (NAR) paradigms (e.g., E3-TTS~\citep{gao2023e3ttseasyendtoend}, F5-TTS~\citep{chen2024f5}, F5R-TTS~\citep{sun2025f5rttsimprovingflowmatchingbased}, et al.); (3) hybrid AR-NAR architectures (e.g., CosyVoice~\citep{du2024cosyvoice}, FireRedTTS~\citep{guo2024fireredtts}, Seed-TTS\citep{anastassiou2024seed}, MiniMax-Speech~\citep{zhang2025minimax}, and IndexTTS2~\citep{zhou2025indextts2}, et al.), which have collectively narrowed the gap between synthetic and human speech.

Despite these achievements, recent state-of-the-art (SOTA) TTS systems still face critical challenges that hinder production-level deployment. First, high-quality voice cloning typically requires large-scale training data and relatively long reference recordings, limiting applicability in low-resource scenarios. Second, emotional expressiveness remains constrained. Most models either fail to capture nuanced text-related emotions or rely on explicit emotion labels that complicate workflows and generalization. Third, the precision of pronunciation for polyphonic characters, rare words, and dialects remains suboptimal, especially in languages such as Chinese with rich phonetic variations. Fourth, reinforcement learning (RL), while promising for aligning speech outputs with human preferences, is underexplored in TTS due to difficulties in reward design and training stability. Finally, adapting premium or personalized voices often relies on costly full-model fine-tuning, making scalable customization impractical in production settings.

To address these limitations, we present GLM-TTS, a production-level TTS system optimized for efficiency, controllability, and naturalness. As shown in Figure~\ref{img:GLM_TTS_overall}, built on a two-stage generation paradigm inspired by CosyVoice (Text-to-Token Autoregressive + Token-to-Wav Diffusion), GLM-TTS achieves SOTA performance on open-source benchmarks with only 100k hours of training data, which is far less than large-scale counterparts like CosyVoice 3 (1M hours)~\citep{du2025cosyvoice3} and FireRedTTS-2 (1.1M hours)~\citep{xie2025fireredtts2}.

Our contributions are structured around the practical requirements of industrial TTS deployment:
\begin{itemize}  
 \item[$\bullet$] Speech Tokenizer: Leveraging an optimized Whisper-VQ speech tokenizer with fundamental frequency constraints and expanded vocabulary (32k), GLM-TTS achieves high speaker similarity (SIM = 76.1) and low character error rate (CER = 1.03\%) on Seed-TTS-eval zh test-set.
  \item[$\bullet$]Multi-Reward Reinforcement Learning: Adopting a GRPO-based RL framework, we fuse four critical rewards (CER for pronunciation accuracy, SIM for timbre fidelity, Emotion for expressive naturalness, and Laughter for paralinguistic realism) with dynamic sampling and gradient clipping. This resolves the reward hacking and training instability issues plaguing prior RL-based TTS, enabling GLM-TTS to outperform commercial models in nuanced emotion expression (e.g., happy, sadness, and anger) on the CV3-eval-emotion benchmark and to achieve superior WER and SIM metrics on the Seed-TTS-eval benchmark as well.
 \item[$\bullet$]Low-Cost Premium Voice Customization: Through optimized LoRA fine-tuning, GLM-TTS achieves full-model-level performance by adjusting only 15\% of parameters—reducing data requirements to 1 hour of single-speaker audio and training costs by 80\% compared to full fine-tuning.
\item[$\bullet$]Precision Pronunciation Control: A ``Hybrid Phoneme + Text'' input scheme with a dynamically extensible dictionary addresses polyphonic and rare word ambiguities, a longstanding challenge in Chinese TTS, without sacrificing prosodic naturalness.
\item[$\bullet$]Enhanced Waveform Reconstruction: A novel Vocos2D vocoder replaces 1D convolutions with 2D operations and DiT-style residual connections, improving frequency subband modeling. Mixed training with high-quality singing data expands the vocal range and adapts the model to complex acoustic conditions.
\end{itemize}

\section{Methodology}
\subsection{Data processing pipeline}
Leveraging proprietary audio datasets, we have constructed a comprehensive and robust data processing pipeline to generate high-quality audio data for subsequent model training. The data pipeline consists of the following steps:
\begin{figure}[h]
\begin{center}
\includegraphics[width=.6\textwidth]{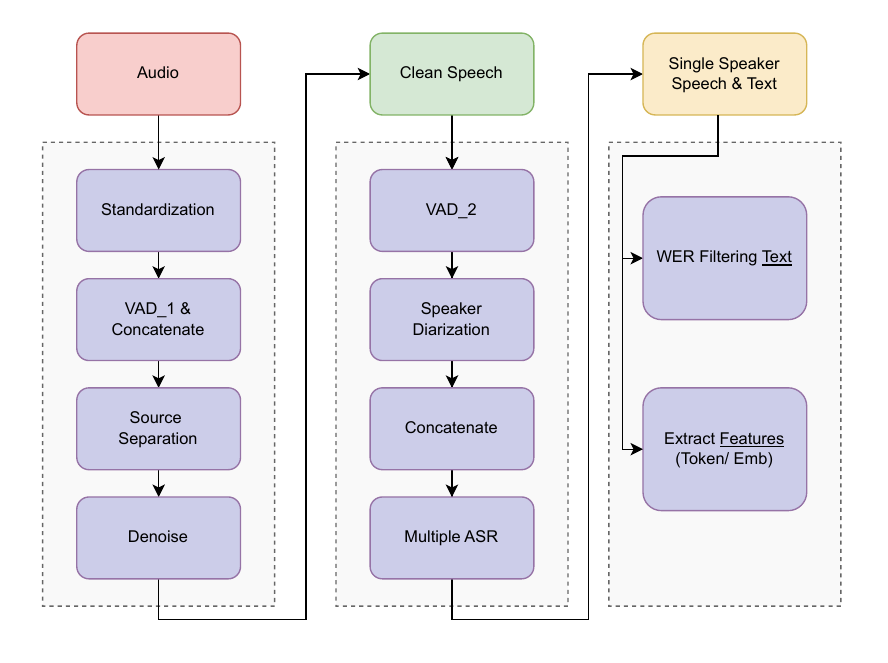}
\end{center}
\caption{An overview of the data processing pipeline.}
\end{figure}

\begin{itemize}
    \item \textbf{Speech Standardization and Coarse Segmentation}.
    First, we unify heterogeneous audio data into the WAV format to eliminate format-related inconsistencies. Then, Voice Activity Detection (VAD) technology~\citep{pyannote} is applied to segment valid speech fragments from the original audio, and these valid fragments are then concatenated into long audio clips of approximately 10 minutes for subsequent processing steps.
    
    \item \textbf{Source Separation and Denoising}.
    To obtain clean speech signals, we first adopt the Mel-Band Roformer model (an improved variant of the RoFormer model~\citep{roformer}) to separate background sounds from the speech signals. After that, a self-developed denoising model is used to further suppress residual noise, ensuring the purity of the speech data.
    
    \item \textbf{Speaker Diarization and Concatenation}.
    We leverage the pyannote.audio model~\citep{pyannote} to implement multi-speaker separation, which can accurately distinguish speech fragments from different speakers. For the speech fragments of a single speaker, we perform amplitude normalization to ensure consistent volume levels, and then concatenate these fragments until the total length reaches the target duration (capping the sequence length at 40 seconds).
    
    \item \textbf{WER Filtering}.
    To select high-quality audio data, we conduct speech recognition and Word Error Rate (WER) calculation with a double-check mechanism:
    \begin{itemize}
        \item For Chinese audio: Open-source Automatic Speech Recognition (ASR) models including Paraformer~\citep{paraformer} and SenseVoice~\citep{an2024funaudiollmvoiceunderstandinggeneration} are used for transcription.
        \item For English audio: Open-source ASR models including Whisper~\citep{whisper} and Reverb~\citep{reverb} are adopted for transcription.
    \end{itemize}
    We calculate the WER for the transcribed text and retain only the audio data with a WER of less than 5\% to ensure high accuracy of the speech-content correspondence.
    
    \item \textbf{Punctuation Optimization}.
    First, we perform text-speech forced alignment~\citep{alignment} to obtain the pronunciation duration of each character in the transcribed text. Then, we calculate the pronunciation threshold as the sum of the mean and 2.6 times the variance of the character pronunciation durations. Finally, we optimize the punctuation based on the interval between adjacent characters: If the interval exceeds the threshold, we retain or add punctuation. Otherwise, punctuation is omitted.
    
    \item \textbf{Feature Extraction}.
    Based on the filtered single-speaker clean audio data, we extract audio speaker embeddings and speech tokens to support the training of subsequent models.
    
    \item \textbf{Overall Engineering Optimization}.
    To enable large-scale data processing efficiently, we adopt the gRPC framework~\citep{grpc} and a server-worker architecture to accelerate each sub-module of the pipeline. Additionally, we rely on a distributed cluster to fully utilize the memory of multiple GPUs and the batch processing capability, significantly improving the overall processing efficiency of the pipeline.
\end{itemize}

\subsection{Text Tokenizer}
\textbf{Vocabulary Pruning for Alignment Stability}.
To alleviate the modeling burden of semantic-to-acoustic alignment, we prune the tokenizer’s vocabulary by removing tokens composed of more than two Chinese characters. Although we implement heuristic constraints during sampling to bound the speech-to-text length ratio (e.g., within $[2, 20]$), relying solely on such hard constraints is insufficient for optimal convergence. Long text tokens inherently exhibit high variance in acoustic duration and frequently stretch the upper bounds of the ratio, creating sparse and difficult-to-learn mappings. By enforcing a finer text granularity, we intrinsically normalize the information density and center the length ratio distribution. This structural optimization effectively alleviates the burden of learning extreme one-to-many text-to-acoustic alignments, ensuring robust autoregressive generation even at a high speech token rate of 25 Hz.

\subsection{Speech Tokenizer}
GLM-TTS introduces a series of optimizations to the Whisper-VQ speech tokenizer~\citep{whisper} based on GLM-4-Voice~\citep{glm4voice}, aimed at improving pronunciation accuracy, naturalness, and expressiveness:
\begin{itemize}
    \item \textbf{Increased Token Generation Rate}. The token rate is doubled from 12.5Hz to 25Hz, and the vocabulary size is expanded from 16k to 32k. This enhancement effectively reduces pronunciation glitches at high speaking speeds and improves the naturalness of paralinguistic features such as laughter and breathing sounds.
    \item \textbf{Introduction of Pitch Estimator (PE) Module}. A new pitch estimation module has been added to optimize pitch modeling accuracy, thereby improving the prosody alignment between the cloned TTS output and reference (prompt) audio.
    \item \textbf{Adoption of Non-Causal Architecture}. The original causal constraint has been lifted. The block attention structure was removed, and causal convolution has been replaced with standard convolution, removing sequential bottlenecks and improving the accuracy of both the ASR and PE modules.
    \item \textbf{Expanded Training Data Scope}. The scale and diversity of training data are significantly increased. Large-scale dialect datasets have been incorporated to strengthen dialect comprehension, and high-quality singing voice data have been added to enrich the model’s phonetic learning samples, further enhancing adaptability to diverse scenarios.
\end{itemize}

\subsection{SpeechLM RL-Alignment}
\begin{figure*}[h]
\begin{center}
\includegraphics[width=\textwidth]{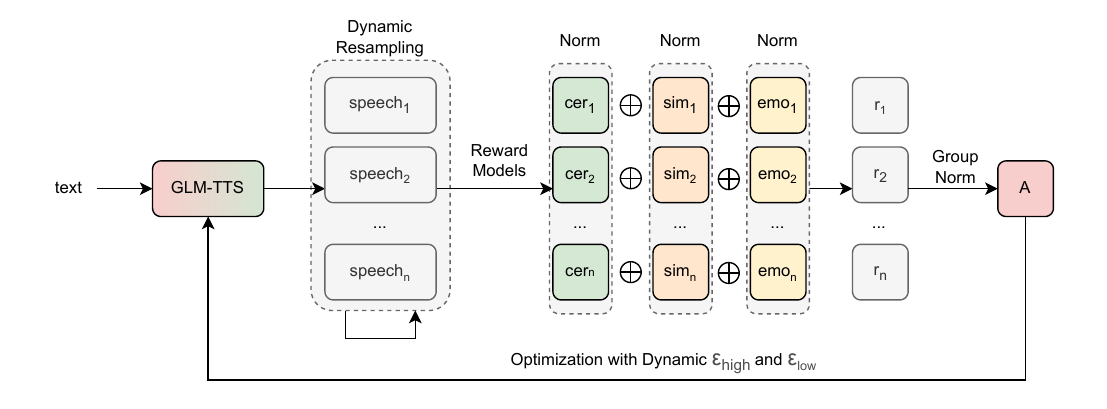}
\end{center}
\caption{An overview of the GLM-TTS-GRPO framework.}
\label{fig:rl}
\end{figure*}

Reinforcement learning has not yet been widely applied in speech synthesis, with major bottlenecks stemming from the complexity of reward mechanism design and the propensity for gradient vanishing or performance degradation during training. GLM-TTS addresses these challenges by introducing the GRPO~\citep{shao2024deepseekmathpushinglimitsmathematical} reinforcement learning paradigm along with a series of strategies, significantly enhancing the core capabilities of both pre-trained and SFT models—including pronunciation accuracy, timbre similarity, and overall naturalness. Furthermore, GLM-TTS achieves superior human-like qualities, notably in emotional expressivity and the naturalness of paralinguistic features. Figure~\ref{fig:rl} illustrates the GLM-TTS-GRPO framework.
Leveraging the GRPO algorithmic framework, we achieve significant performance improvements through three major design advancements:

First, we introduce a multidimensional regularization reward mechanism, integrating four core rewards—CER (character error rate), SIM (similarity), Emotion, and Laughter. By employing a hierarchical processing strategy (``individual reward regularization → weighted fusion → overall regularization''), we effectively address the issue of different reward distributions.

Second, we implement a dynamic sampling strategy to mitigate potential gradient vanishing within a batch. This mechanism automatically triggers resampling (up to three times) when batch rewards become homogeneous, while limiting the total number of resampling to avoid negative optimization from poor-quality samples, thereby balancing training stability and efficiency.

Third, we adopt an adaptive gradient clipping scheme, setting $\epsilon_{high}$ and $\epsilon_{low}$ as dynamic values that adjust according to training steps. In the early stages, tighter clipping prevents the model from quickly exploiting reward hacking shortcuts; in the later stages, the constraints are gradually relaxed to allow for broader exploration. Reasonable parameter ranges ensure the number of clipped tokens remains stable, preventing ineffective clipping or excessive restriction. Additionally, setting $\epsilon_{high} > \epsilon_{low}$ encourages the generation of low-probability tokens, significantly enhancing the human-likeness of synthesized speech.

\subsection{LoRA for Premium Voice Customization}

Full parameter fine-tuning in large speech models is often costly and unstable due to uneven data quality. To address this, we optimize the LoRA (Low-Rank Adaptation) fine-tuning paradigm for ``Premium Voice Customization''. This approach aims to achieve stable, high-quality voice customization with low resource and cost overhead.

\begin{itemize}
    \item \textbf{Efficiency and Efficacy}. By fine-tuning only about 15\% of the core backbone parameters for approximately 100 epochs, we achieve voice similarity and naturalness comparable to full parameter fine-tuning. This is a significant improvement over initial explorations where tuning only $0.3\%-5\%$ of parameters resulted in limited improvement in voice style and emotion.
    \item \textbf{Cost Efficiency and Data Robustness}. Customization requires only about 1 hour of high-quality single-speaker audio, significantly lowering development costs and barriers. This streamlined process eliminates the need for large-batch data testing and complex data matching and quality filtering, which are often problematic due to the high inconsistency in data distribution and quality in SFT.
    \item \textbf{Stability}. Controlling the ratio of fine-tuned parameters (e.g., above 15\%) enhances generalization and stability across different scenarios, ensuring production-level reliability. The improved stability is crucial as complex requirements for small-batch, high-demand voice customization are difficult to implement effectively using full parameter fine-tuning.
\end{itemize}

\subsection{Phoneme-in}
In professional speech synthesis scenarios, such as education and standardized testing, there is an exceptionally high demand for pronunciation accuracy. Traditional large-scale TTS models typically rely on automatic sampling or default probabilities when handling complex linguistic features like polyphones (characters with multiple pronunciations) and rare characters. This lack of explicit control mechanisms often results in uncontrollable pronunciation and higher error rates. To address this, we introduce \textbf{Phoneme-in}, an enhancement capability that utilizes phoneme-level input to achieve precise, controllable pronunciation.

\textbf{Vocabulary Construction and Regularization.} We construct dedicated vocabularies for polyphones and rare characters to aggregate terms requiring precise control in key application scenarios. These vocabularies support the downstream logic for targeted phoneme replacement and allow for dynamic maintenance and expansion based on specific business requirements.

\textbf{Hybrid Training Paradigm.} To equip the model with the ability to understand and adapt to phoneme inputs, we employ a mixed-modality training strategy:
\begin{itemize}
    \item For standard characters (excluding defined polyphones and rare characters), we employ a \textbf{two-stage probabilistic replacement strategy}. 
    During training, the replacement process is triggered with a specific probability (e.g., $p=0.2$). 
    Once triggered, a random subset of characters is converted into phonemes, where the replacement ratio is uniformly sampled between $0$ and a maximum threshold (e.g., $0.5$). 
    This dynamic ``Hybrid Phoneme + Text'' augmentation significantly enhances the model's robustness to mixed-modality inputs.
    \item Characters belonging to the polyphone or rare character vocabularies are preserved as original text without conversion during training, ensuring the model retains semantic context for these complex cases.
\end{itemize}

\textbf{Fine-grained Control at Inference.} The inference process is designed to maximize precision:
\begin{enumerate}
    \item The system first processes the entire input sentence through a Grapheme-to-Phoneme (G2P) module to generate a complete phoneme sequence (phoneme\_list).
    \item It iterates through the original text list; if a polyphone or rare character is encountered, the text is replaced with its corresponding phoneme from the generated list.
    \item The final input to the model takes a ``hybrid phoneme+text'' format. This approach ensures precise pronunciation control for ambiguous characters while preserving the natural prosody associated with the standard text.
\end{enumerate}

\subsection{Vocos2D}
\begin{figure}[t]
  \centering
  \includegraphics[page=1,scale=1]{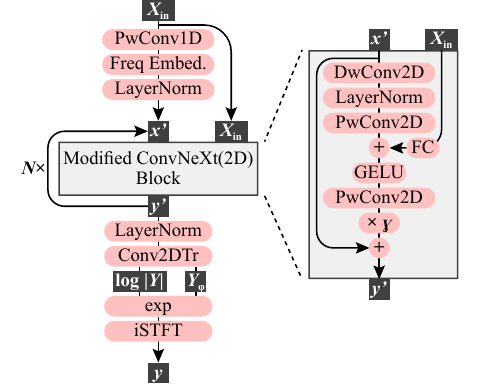}
  \caption{Diagram of the Vocos2D generator architecture. FC denotes a fully connected (linear) layer, PwConv denotes point-wise convolution, and DwConv denotes depth-wise convolution.}
  \label{fig:vocos2d-architecture}
  \vspace{-4mm}
\end{figure}

The original Vocos~\citep{Vocos}, a GAN-based vocoder, uses 1D convolutions in its generator to process entire frames across all frequencies. Drawing inspiration from sub-band processing and image-processing techniques, we redesign the generator to incorporate 2D convolutions, enabling more focused handling of specific frequency subbands. Figure~\ref{fig:vocos2d-architecture} illustrates the Vocos2D generator architecture.

For the input $X_{in}$, an initial point-wise convolution (implemented as a fully connected linear layer) is followed by learned per-frequency embeddings to facilitate inter-frequency communication, as translation invariance does not apply across frequency bins. 

The Vocos2D backbone block adapts the original ConvNeXt design, augmented with additional shortcut connections from the input Mel spectrogram $X_{in}$. These are regressed through a linear layer and added to the inverted bottleneck stage, allowing direct incorporation of the input spectrogram condition.

\begin{figure}[t]
  \centering
  \includegraphics[page=2,scale=1]{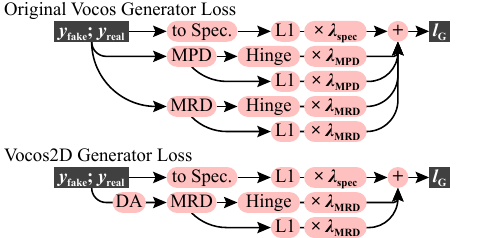}
  \caption{Comparison of the original Vocos generator loss (upper) and the proposed Vocos2D generator loss (lower). DA denotes discriminator augmentation, Hinge denotes hinge loss, MPD denotes multi-period discriminator, and MRD denotes multi-resolution discriminator.}
  \label{fig:vocos2d-generator-loss}
  \vspace{-4mm}
\end{figure}

\textbf{Generator Loss.} Figure~\ref{fig:vocos2d-generator-loss} compares the generator losses of the original Vocos and Vocos2D. We make two key changes: (1) removing the multi-period discriminator, as it degraded performance on input linear spectrograms with more frequency bins; and (2) adding discriminator augmentation (DA)~\citep{DiffAugment,ADA} before feeding real and fake waveforms into the multi-resolution discriminator~\citep{UnivNet}. DA improves training stability by applying differentiable transformations only to the discriminator, allowing gradients to flow back to the generator without forcing it to model augmentations. We use three transformations: random loudness adjustments within $\pm 6$ dB, random sample shifts, and random phase rotations.

\textbf{Discriminator Training.} Apart from the removal of the multi-period discriminator, Vocos2D's discriminator training mirrors the original Vocos, using hinge loss~\citep{SAGAN,TEGAN} and the multi-resolution discriminator~\citep{UnivNet}.

\textbf{Dataset.} To support 32 kHz high-quality wideband speech synthesis, we augment our proprietary speech dataset with high-quality open-source singing voice data, expanding pitch range coverage and enhancing overall sound quality and adaptability to varied vocalization techniques.

\section{Experiments and Results}

\subsection{Speech tokenizer Evaluation}
We evaluate GLM-TTS-tokenizer across two English test sets and five Chinese dialect and accented Mandarin test sets, comparing with GLM4-Voice-tokenizer. As shown in Table~\ref{tab:tokenizer_ASR}, the GLM-TTS-tokenizer significantly outperforms the GLM-4-Voice tokenizer in Chinese recognition, and shows minor improvements in English recognition.

Further, we implement TTS systems on both tokenizers and evaluate TTS metrics. The results in Table~\ref{tab:tokenizer_TTS} show that GLM-TTS outperforms the variant built on GLM4-Voice-tokenizer in terms of Speaker Similarity(SIM) and CER.

\begin{table}[h]

\centering
\small
\setlength{\tabcolsep}{5pt}
\renewcommand{\arraystretch}{1.15}
\caption{Performance of TTS systems built on GLM4-Voice-tokenizer and GLM-TTS-tokenizer on ASR tasks. Values represent WER/CER, the lower, the better.}
\begin{tabular}{lccccccc}
\toprule
\multirow{2}{*}{\textbf{Tokenizer}} & \textbf{Libri} & \textbf{Libri} & \textbf{Sichuan} & \textbf{Jiao-Liao} & \textbf{Taiwan} & \multirow{2}{*}{\textbf{Cantonese}}& \textbf{Shanghai} \\
& \textbf{Other} & \textbf{Clean} & \textbf{dialect} & \textbf{Mandarin} & \textbf{Mandarin} & & \textbf{dialect}\\
\midrule
GLM4-Voice-tokenizer & 4.90  &  \best{2.10} & 54.11 & 14.04   & 49.09 & 46.81 & 72.06  \\
GLM-TTS-tokenizer & \best{4.51}   &  2.12   & \best{24.40} & \best{9.11}   & \best{16.92} & \best{7.27} & \best{19.15}  \\
\bottomrule
\end{tabular}
\label{tab:tokenizer_ASR}

\end{table}

\begin{table}[h]
\centering
\small
\setlength{\tabcolsep}{8pt}
\renewcommand{\arraystretch}{1.15}
\caption{Performance of GLM4-Voice-tokenizer and GLM-TTS-tokenizer on seed\_test\_zh.}
\begin{tabular}{lcc}
\toprule
\textbf{Tokenizer} & \textbf{SIM}$\uparrow$ & \textbf{CER}$\downarrow$ \\
\midrule
GLM4-Voice-tokenizer & 75.2  &  1.44 \\
GLM-TTS-tokenizer & \best{76.1}   &  \best{1.03} \\
\bottomrule
\end{tabular}
\label{tab:tokenizer_TTS}
\vspace{-2mm}
\end{table}

\begin{table}[t]
\centering
\caption{Performance comparison on different test sets. \textbf{test-zh}: Chinese standard test set; \textbf{test-en}: English test set; \textbf{test-hard}: Hard case test set (polyphones, rare words). CER/WER are lower-is-better ($\downarrow$), SIM is higher-is-better ($\uparrow$).}
\label{tab:main-results}
\small
\setlength{\tabcolsep}{2.5pt}
\renewcommand{\arraystretch}{1.15}

\begin{tabular}{lcc cc cc}
\toprule
\multirow{2}{*}{\textbf{Model}} & \multirow{2}{*}{\textbf{Params}} & \multirow{2}{*}{\textbf{\parbox{1.3cm}{\centering Open\\source}}} & \multicolumn{2}{c}{\textbf{\textit{test-zh}}} & \multicolumn{2}{c}{\textbf{\textit{test-en}}}  \\
\cmidrule(lr){4-5} \cmidrule(lr){6-7} 
& & & \textbf{CER} $\downarrow$ & \textbf{SIM} $\uparrow$ & \textbf{WER} $\downarrow$ & \textbf{SIM} $\uparrow$ \\
\midrule
MegaTTS3~\citep{jiang2025megatts3} & 0.5B & No & 1.52 & 79.0 & 2.79 & 77.1 \\
DiTAR~\citep{jia2025ditar} & 0.6B & No & 1.02 & 75.3 & 1.69 & 73.5  \\
CosyVoice3~\citep{du2025cosyvoice3} & 1.5B & No & 1.12 & 78.1 & 2.22 & 72.0  \\
Seed-TTS~\citep{anastassiou2024seed} & - & No & 1.12 & \best{79.6} & 2.25 & \best{76.2} \\
MiniMax-Speech~\citep{zhang2025minimax} & - & No & \best{0.83} & 78.3 & \best{1.65} & 69.2  \\
\midrule
F5-TTS~\citep{chen2024f5} & 0.3B & Yes & 1.53 & 76.0 & 2.00 & 67.0  \\
MaskGCT~\citep{wang2024maskgct} & - & Yes & 2.27 & 77.4 & 2.62 & 71.7  \\
CosyVoice~\citep{du2024cosyvoice} & 0.3B & Yes & 3.63 & 72.3 & 4.29 & 60.9  \\
CosyVoice2~\citep{du2024cosyvoice2} & 0.5B & Yes & 1.38 & 75.7 & 3.09 & 65.9  \\
CosyVoice3~\citep{du2025cosyvoice3} & 0.5B & Yes & 1.16 & \best{78.0} & 2.02 & 71.8  \\
SparkTTS~\citep{wang2025spark} & 0.5B & Yes & 1.54 & 66.0 & 3.14 & 57.3  \\
FireRedTTS~\citep{guo2024fireredtts} & 0.5B & Yes & 1.51 & 63.5 & 3.82 & 46.0  \\
FireRedTTS-2~\citep{xie2025fireredtts2} & - & Yes & 1.14 & 73.6 & 1.95 & 66.5  \\
Qwen2.5-Omni~\citep{xu2025qwen2} & 7B & Yes & 1.70 & 75.2 & 2.72 & 63.2  \\
OpenAudio-s1-mini~\citep{openaudio2024openaudio} & 0.5B & Yes & 1.18 & 68.5 & 1.94 & 55.0  \\
IndexTTS 2~\citep{zhou2025indextts2} & 1.5B & Yes & 1.03 & 76.5 & 2.23 & 70.6  \\
VibeVoice~\citep{peng2025vibevoice} & 1.5B & Yes & 1.16 & 74.4 & 3.04 & 68.9  \\
HiggsAudio-v2~\citep{bosonai2025higgs} & 3B & Yes & 1.50 & 74.0 & 2.44 & 67.7  \\
VoxCPM~\citep{zhou2025voxcpmtokenizerfreettscontextaware} & 0.5B & Yes & 0.93 & 77.2 & \best{1.85} & \best{72.9}  \\
\midrule
\textbf{GLM-TTS (Ours)} & 1.5B & Yes & 1.03 & 76.1 & 2.23 & 67.2  \\
\textbf{GLM-TTS\_RL (Ours)} & 1.5B & Yes & \best{0.89} & 76.4 & 1.91 & 68.1  \\
\bottomrule
\end{tabular}
\end{table}

\subsection{Voice Cloning Results on Seed-TTS-eval}

Table~\ref{tab:main-results} reports results on the Seed-TTS-eval benchmark using standard TTS metrics: Character Error Rate (CER) and Word Error Rate (WER) for pronunciation accuracy (lower is better), and Speaker Similarity (SIM, higher is better) measured by calculating the cosine similarity between speaker embeddings extracted using fine-tuned WavLM-large\citep{chen2022largescaleselfsupervisedspeechrepresentation}. We compare GLM-TTS (1.5B parameters, open-source) with 19 state-of-the-art baselines, including both closed-source systems (e.g., MiniMax-Speech, Seed-TTS, et al.) and open-source models (e.g., IndexTTS2, FireRedTTS-2, VibeVoice, et al.).

On the \texttt{test-zh} set, closed-source models deliver leading performance: MiniMax-Speech achieves a state-of-the-art CER of 0.83\%, while Seed-TTS attains a standout SIM of 79.6. GLM-TTS attains a CER of 1.03\% and SIM of 76.1, delivering results that are generally comparable to the open-source SOTA TTS model, such as VoxCPM and IndexTTS2. After applying our multi-reward GRPO reinforcement learning, GLM-TTS\_RL further improves to CER=0.89\% and SIM=76.4, substantially narrowing the gap with the leading closed-source systems.

\textbf{Notably, due to practical industrial constraints, GLM-TTS is trained with a limited amount of English data (less than half of the Chinese training data).}
Despite this limitation, on the \texttt{test-en} set, GLM-TTS still achieves reasonable WER (2.23\%) and SIM (67.2) performance on the English test set, suggesting potential for further improvement with additional English data.

Overall, GLM-TTS achieves top-tier performance among 1.5B-scale open-source models on the Seed-TTS-eval test set, with a negligible performance gap relative to state-of-the-art closed-source counterparts. These results demonstrate the effectiveness of the proposed overall framework, and further validate the effectiveness of our core design choices—multi-reward GRPO reinforcement learning for consistently improving pronunciation accuracy and timbre fidelity.

\subsection{Speech-LM RL-Alignment} 
We adopt two techniques from DAPO~\citep{yu2025dapoopensourcellmreinforcement}: \textbf{Clip-Higher} and \textbf{Dynamic Sampling}. First, we set $\epsilon_{high}$ to 0.3 and $\epsilon_{low}$ to 0.2. For Dynamic Sampling, to facilitate training, we resample batches with \textit{zero} advantages up to three times. As shown in Table~\ref{tab:ablation_cd}, Clip-Higher yields fully positive gains. Dynamic Sampling improves both SIM and CER, yet leads to a decline in EMO. This is because repeated sampling can easily induce variance in SIM or CER across different samples, while the distribution of EMO is more extreme, approximating a bimodal distribution concentrated near 0 and 1. Consequently, we no longer consider the variance introduced by SIM in the resampling process.

\begin{table}[h]

\centering
\small
\setlength{\tabcolsep}{10pt}
\renewcommand{\arraystretch}{1.15}
\caption{Performance of Pretrain-GRPO with Clip-Higher and Dynamic Sampling on an internal emotion test set. $c$ represents Clip-Higher and $d$ represents Dynamic Sampling. }
\begin{tabular}{lccccccc}
\toprule
\textbf{Model} & \textbf{CER}$\downarrow$ & \textbf{SIM}$\uparrow$ & \textbf{EMO}$\uparrow$ \\
\midrule
Pretrain-base & 2.05  & 80.0 & 0.525  \\
Pretrain-GRPO & 1.99 &  80.3  & 0.565   \\
Pretrain-GRPO\_c & 1.93 &  80.4  & 0.660   \\
Pretrain-GRPO\_d & 1.91 &  80.8  & 0.440   \\
\bottomrule
\end{tabular}
\label{tab:ablation_cd}
\end{table}

Since Clip-Higher has demonstrated improvement, we further explored dynamically adjusting parameters such as $\epsilon_{h}$ during training to provide the model with sufficient exploration space while maintaining training stability. Specifically, we selected three parameters, $\epsilon_h$, $\epsilon_l$, and $T$ with initial values of 0.3, 0.2, and 1, respectively, and allowed them to linearly increase as training progressed until training completion. As shown in Table~\ref{tab:RL_dynamic_p}, we observed that more aggressive parameter settings tend to lead to less stable training: while emotional expressiveness is enhanced, pronunciation becomes less clear. Moreover, excessively loose constraints are also prone to resulting in reward hacking.
\begin{table}[h]

\centering
\small
\setlength{\tabcolsep}{10pt}
\renewcommand{\arraystretch}{1.15}
\caption{Performance of SFT-GRPO with dynamic $\epsilon_h$, $\epsilon_l$, and $T$ on an internal emotion test set. $*$ means freezing parameters during training process.}
\begin{tabular}{lccccccc}
\toprule
\multirow{2}{*}{\textbf{Model}} & \multicolumn{3}{c}{\textbf{\textit{params}}} & \multicolumn{3}{c}{\textbf{\textit{metrics}}} \\
& T & $\epsilon_{h}$ & $\epsilon_{l}$ &
\textbf{CER}$\downarrow$ & \textbf{SIM}$\uparrow$ & \textbf{EMO}$\uparrow$ \\
\midrule
SFT-base & -& -& - & 2.13  & 76.1 & 0.695  \\
SFT-GRPO$^*$ & - &- & - & 2.21 &  76.3  & 0.720   \\
SFT-GRPO & 1.5 & 0.5 & 0.4 & 2.09 &  76.7  & 0.705   \\
SFT-GRPO & 2 & 1 & 0.4 & 2.16 & 75.5 & 0.790   \\
SFT-GRPO & 3 & 1 & 0.4 & 2.21 & 78.1 & 0.885   \\
\bottomrule
\end{tabular}
\label{tab:RL_dynamic_p}

\end{table}

To further improve laughter modeling, we introduced a laughter reward, which can be summarized as follows: if the text contains two or more consecutive laughter words (such as ``ha'' and ``hey''), and the laughter detection model identifies a laughter segment, then (1) if the ASR system transcribes the segment as a ``deletion''(empty string), the reward is set to 1; (2) if the ASR system transcribes the corresponding text, the reward is set to 0.

We experimented with different weights $\lambda_{laughter}$ for the laughter reward and the results are shown in Table~\ref{tab:RL_laughter}. Increasing the laughter reward leads to decreases in CER and similarity scores, as laughter segments cannot be recognized by the ASR model as textual content, and the timbre of laughter differs from the speaker's regular speaking voice. Nevertheless, enhancing laughter synthesis can also improve the model's emotional expressiveness.

\begin{table}[h]

\centering
\small
\setlength{\tabcolsep}{10pt}
\renewcommand{\arraystretch}{1.15}
\caption{Performance of SFT-GRPO with different $\lambda_{laughter}$ on an internal emotion test set.}
\begin{tabular}{lccccccc}
\toprule
\multirow{2}{*}{\textbf{Model}} & \multicolumn{1}{c}{\textbf{\textit{params}}} & \multicolumn{3}{c}{\textbf{\textit{metrics}}} \\
& $\lambda_{laughter}$ &
\textbf{CER}$\downarrow$ & \textbf{SIM}$\uparrow$ & \textbf{EMO}$\uparrow$ \\
\midrule
SFT-base & - & 3.11  & 76.3 & 0.44 \\
SFT-GRPO & 2 & 2.86 &  74.6  & 0.64 \\
SFT-GRPO & 5  & 3.18 &  74.8  & 0.66   \\
SFT-GRPO & 10  & 3.06 &  74.8  & 0.72   \\
\bottomrule
\end{tabular}
\label{tab:RL_laughter}

\end{table}

\subsection{Effectiveness of Phoneme-in}

To rigorously evaluate the impact of the \textbf{Phoneme-in} mechanism on pronunciation accuracy, we conducted a targeted ablation study using a proprietary internal dataset. Unlike standard public benchmarks, this dataset is specifically constructed to simulate challenging industrial scenarios, characterized by a high density of polyphonic characters, low-frequency words, and ambiguous contexts that typically confuse end-to-end TTS systems.

We compared the performance of GLM-TTS with and without the Phoneme-in module enabled. As illustrated in Table~\ref{tab:phoneme_in_ablation}, the baseline model, which relies solely on text input and implicit grapheme-to-phoneme prediction, yields a Phoneme Error Rate (PER) of 13.23\%. This relatively high error rate indicates the inherent difficulty of the test set. However, when the Phoneme-in mechanism is activated—allowing for fine-grained, phoneme-level intervention—the PER dramatically drops to 5.14\%. 


\begin{table}[h]
\centering
\small
\setlength{\tabcolsep}{10pt}
\renewcommand{\arraystretch}{1.15}
\caption{Ablation study of the Phoneme-in mechanism on the internal hard-case dataset. The use of hybrid phoneme input significantly reduces pronunciation errors.}
\begin{tabular}{lccc}
\toprule
\textbf{Model Settings} & \textbf{Input Modality} & \textbf{PER} ($\downarrow$) \\
\midrule
GLM-TTS (w/o Phoneme-in) & Text Only & 13.23 \\
GLM-TTS (w/ Phoneme-in) & Hybrid (Text + Phoneme) & \best{5.14} \\
\bottomrule
\end{tabular}
\label{tab:phoneme_in_ablation}
\end{table}

\subsection{Vocos2D vocoder} 
\begin{table}[H]
  \centering
  \small
  \setlength{\tabcolsep}{8pt}
  \renewcommand{\arraystretch}{1.15}
  \caption{Performance comparison between Vocos and Vocos2D.}
  \begin{tabular}{rcccc}
  \toprule
    & \textbf{NISQA}$\uparrow$ & \textbf{UTMOS}$\uparrow$ & \textbf{Ab. Aes.-PQ}$\uparrow$ & \textbf{MOS}$\uparrow$ \\
  \midrule
  GT      & 3.47 & 2.11 & 7.68 & 4.77 \\
  \midrule
  Vocos   & 3.16 & 1.87 & 7.56 & 3.58 \\
  Vocos2D & 3.40 & 1.91 & 7.64 & 4.16 \\
  \bottomrule
  \end{tabular}
  \label{tab:vocos2d-results}
\end{table}

We evaluate Vocos2D on an internal test set of randomly selected audio samples. Objective metrics include UTMOS~\citep{UTMOS}, NISQA~\citep{NISQA}, and the production quality (PQ) score from Meta AudioBox Aesthetics~\citep{AudioboxAesthetics}, complemented by subjective MOS evaluations. As shown in Table~\ref{tab:vocos2d-results}, Vocos2D consistently outperforms the original Vocos across all metrics, demonstrating the effectiveness of the proposed architectural and training improvements.

\section{Conclusion}

In this technical report, we introduce \textbf{GLM-TTS}, a production-level text-to-speech system that systematically addresses several long-standing challenges in modern TTS. With only 100k hours of training data, GLM-TTS achieves competitive pronunciation accuracy and speaker similarity. The proposed multi-reward GRPO-based reinforcement learning framework effectively aligns synthesized speech with human perceptual preferences, leading to consistent improvements in pronunciation accuracy, emotional expressiveness, and speaker similarity without sacrificing training stability.

Beyond core model performance, GLM-TTS emphasizes practical deployability. The optimized LoRA-based premium voice customization strategy significantly reduces both data and computational costs, enabling scalable personalization in production environments. The hybrid phoneme-text input mechanism provides precise and controllable pronunciation for polyphonic and rare words, addressing a critical requirement in professional and educational TTS applications, especially for Chinese. Overall, GLM-TTS provides a practical framework for efficient and controllable speech synthesis. We hope it can serve as a foundation for future research on expressive, customizable, and scalable speech generation.

\bibliography{iclr2026_conference}
\bibliographystyle{iclr2026_conference}

\end{document}